\documentclass{PoS}
\usepackage{amsmath,amssymb,amsfonts,graphics,graphicx,amscd,amsfonts,epsfig,color}
\newcommand{\rf}       [1]{(\ref{#1})}
\newcommand{\vev}      [1]{{\ensuremath {\langle {#1} \rangle}}}
\newcommand{\mf}          {\ensuremath{m_{\rm f}}}
\newcommand{\epsilonm}    {\ensuremath{\epsilon}}
\newcommand{\Sb}          {\ensuremath{S_{\mathrm{b}}}} 
 
\newcommand{\Seff}        {\ensuremath{S_{\mathrm{eff}}}}

\newcommand {\beq} {\begin{equation}}
\newcommand {\eeq} {\end{equation}}
\newcommand {\beqa}{\begin{eqnarray}}
\newcommand {\eeqa}{\end{eqnarray}}

\newcommand {\tr}{{\textrm{tr}\,}}
\newcommand {\Tr}{\mbox{Tr\,}}

\newcommand {\Pf}{\mbox{Pf}}

\title{Dynamical Compactification of Extra Dimensions in the Euclidean IKKT Matrix Model via Spontaneous Symmetry Breaking}

\ShortTitle{Dynamical Compactification of Extra Dimensions}

\author{Konstantinos N. Anagnostopoulos\\
Physics Department, National Technical University of Athens, Zografou Campus, GR-15780 Zografou, Greece\\
E-mail: \email{konstant@mail.ntua.gr}}

\author{Takehiro Azuma\\
Institute for Fundamental Sciences, Setsunan University, 17-8 Ikeda Nakamachi, Neyagawa, Osaka, 572-8508, Japan\\
E-mail: \email{azuma@mpg.setsunan.ac.jp}}

\author{Yuta Ito\\
National Institute of Technology, Tokuyama College, Gakuendai, Shunan, Yamaguchi 745-8585, Japan
and
KEK Theory Center, High Energy Accelerator Research Organization, 1-1 Oho, Tsukuba, Ibaraki 305-0801, Japan\\
E-mail: \email{y-itou@tokuyama.ac.jp}}

\author{Jun Nishimura\\
KEK Theory Center, High Energy Accelerator Research Organization, 1-1 Oho, Tsukuba, Ibaraki 305-0801, Japan
and
Graduate University for Advanced Studies (SOKENDAI), 1-1 Oho, Tsukuba, Ibaraki 305-0801, Japan\\
E-mail: \email{jnishi@post.kek.jp}}

\author{Toshiyuki Okubo\\
Faculty of Science and Technology, Meijo University, Nagoya, 468-8502, Japan\\
E-mail: \email{tokubo@meijo-u.ac.jp}}

\author{\speaker{Stratos Kovalkov Papadoudis}\\
Physics Department, National Technical University of Athens, Zografou Campus, GR-15780 Zografou, Greece\\
E-mail: \email{sp10018@central.ntua.gr}}

\abstract{The IKKT matrix model has been conjectured to provide a
  promising nonperturbative formulation of superstring theory.  In
  this model, spacetime emerges dynamically from the microscopic
  matrix degrees of freedom in the large-$N$ limit, and Monte Carlo
  simulations of the Lorentzian version provide evidence of an
  emergent (3+1)-dimensional expanding space-time.  In this talk, we
  discuss the Euclidean version of the IKKT matrix model and provide
  evidence of dynamical compactification of the extra dimensions via
  the spontaneous symmetry breaking (SSB) of the 10D rotational
  symmetry. We perform numerical simulations of a system with a severe
  complex action problem by using the complex Langevin method
  (CLM). The CLM suffers from the singular-drift problem and we deform
  the model in order to avoid it. We study the SSB pattern as we vary
  the deformation parameter and we conclude that the original model
  has an SO($3$) symmetric vacuum, in agreement with previous
  calculations using the Gaussian expansion method (GEM). We employ
  the GEM to the deformed model and we obtain results consistent with
  the ones obtained by CLM.}

\FullConference{Corfu Summer Institute 2019 "School and Workshops on Elementary Particle Physics and Gravity" (CORFU2019)\\
		31 August - 25 September 2019\\
		Corfu, Greece}
\begin{document}

\section{Introduction}

The type IIB matrix model \cite{Ishibashi:1996xs}, also known as the
IKKT model, has been proposed as a nonperturbative formulation of
superstring theory. Spacetime emerges
dynamically from the eigenvalues of the bosonic matrices in the large-$N$ limit \cite{Aoki:1998vn} and 
it is possible that the extra dimensions are compactified {\it dynamically} via 
a non perturbative mechanism. Furthermore, it is possible that a unique vacuum exists in the
theory, thereby solving the so-called landscape problem.

The action of the model can be formally viewed as the dimensional
reduction of the 10$D$, $\mathcal{N}=1$, SU($N$) super Yang-Mills
(SYM) theory to zero dimensions.  Numerical simulations
\cite{Kim:2011cr,Ito:2017rcr,Ito:2013ywa,Ito:2013qga,Ito:2015mem,Ito:2015mxa,Azuma:2017dcb,
  Aoki:2019tby,Nishimura:2019qal} suggest that {\it continuum} time
emerges dynamically and 3 dimensional space undergoes rapid expansion
after a critical time $t_c$, while the remaining 6 dimensions do not
expand.  The cosmological time is defined by the eigenvalues of the
temporal matrix $A_0$, and their infinite and homogeneous distribution
in the large $N$ limit is a nontrivial dynamic result, allowing one to
define a continuum and infinitely extending time.  Furthermore, the
spatial matrices $A_i$ have a band diagonal structure in the SU($N$)
basis used to diagonalize $A_0$ that makes possible to define space at a
given time. Then one can study the structure of space, and
especially its size, which for $t>t_c$, it is found to be expanding in the three dominant
dimensions. This expansion is exponential at early times  \cite{Ito:2013ywa} and
becomes a power law at later times \cite{Ito:2015mxa}, making possible
the emergence  of a realistic cosmology from the 
dynamics of the microscopic degrees of freedom.

Simulations of the model using Monte Carlo techniques is hard because of the 
complex action problem. This originates from the  ${\rm e}^{i \Sb}$ factor in the path
integral,  where $\Sb$ is the bosonic part of the action. This problem can be avoided
by integrating the scale factor of the bosonic matrices
\cite{Kim:2011cr,Ito:2017rcr,Ito:2013ywa,Ito:2013qga,Ito:2015mem,Ito:2015mxa,Azuma:2017dcb}. 
One obtains a sharply peaked function
of $\Sb$ near the origin, which can then be approximated by a Gaussian. The complex
action problem vanishes, but the approximation favors singular spatial configurations \cite{Aoki:2019tby}.
In \cite{Nishimura:2019qal} the 6D model at late times was studied without this approximation,
and the Complex Langevin Method (CLM) \cite{Parisi:1984cs,Klauder:1983sp} was used in 
order to confront the complex action problem. A two parameter deformation of the model
was used for making the simulations possible, and provided evidence that when
these parameters become small, configurations with  non trivial spatial structure dominate
the path integral. 

In this work we study the Euclidean version of the IKKT matrix
model. This model, as well as related ones, have been studied
extensively
\cite{Hotta:1998en,Austing:2001bd,Austing:2001pk,Ambjorn:2000bf,Ambjorn:2000dx,
  Ambjorn:2001xs,Anagnostopoulos:2001yb,Ito:2016efb,
  Anagnostopoulos:2010ux,Anagnostopoulos:2011cn,Anagnostopoulos:2013xga,
  Anagnostopoulos:2015gua,Anagnostopoulos:2017gos,Anagnostopoulos:2020xai}, because they are
more tractable numerically and are finite
\cite{Austing:2001bd,Austing:2001pk}, making the introduction of
infrared cutoffs unnecessary. There is strong evidence that dynamical
compactification of extra dimensions in these models is realized via
spontaneous symmetry breaking (SSB) of the SO($10$) rotational
symmetry of the model.  The effective action of the model
$\Seff=S_{\rm R}+i\Gamma$, obtained after integrating out the
fermionic degrees of freedom, is complex. The fluctuations of the
phase ${\rm e}^{i \Gamma}$ for a generic SO($d$), $d>3$ symmetric
configuration, suppress $d>3$ configurations, thereby favoring lower
dimensional configurations and triggering the SSB
\cite{Nishimura:2000ds}.  The evidence is provided mainly by studying
the model using the Gaussian Expansion Method (GEM). In
\cite{Nishimura:2001sx,Kawai:2002jk,Aoyama:2006rk,Aoyama:2010ry,Nishimura:2011xy},
it was shown that the SO($3$) vacuum of the IIB matrix model has the
lowest free energy and the ratio between the extended three directions
and the shrunken seven directions is a finite number, larger than one.
Furthermore, the study of phase quenched models, show no SSB of the
SO($10$) rotational symmetry
\cite{Hotta:1998en,Ambjorn:2000bf,Ambjorn:2000dx,Ambjorn:2001xs},
consistent with the expectation that the complex phase ${\rm e}^{i
  \Gamma}$ plays a critical role in suppressing large dimensional
configurations.

The Monte Carlo simulations of the  Euclidean IIB matrix model confront a very strong complex action
problem, which needs to be addressed by using special methods. Straightforward reweighting is
not possible, since it makes the computational effort to increase exponentially with the matrix size $N$, 
which needs to be extrapolated to infinity. A density of states based method was used in
\cite{Anagnostopoulos:2001yb,Ambjorn:2002pz, Ambjorn:2004jk,
  Anagnostopoulos:2010ux,Anagnostopoulos:2011cn,
  Anagnostopoulos:2013xga,Anagnostopoulos:2015gua}, allowing one to study relatively
large systems and providing evidence that SSB occurs from first principles. But it turned out to be hard to determine 
the pattern of SSB, and it was not until recently \cite{Anagnostopoulos:2017gos,Anagnostopoulos:2019ptt,Anagnostopoulos:2020xai} 
that this question could be addressed by using the Complex Langevin Method (CLM)\cite{Parisi:1984cs,Klauder:1983sp,Parisi:1980ys}.
The CLM is applied by complexifying the degrees of freedom and defining a stochastic process where
the expectation values with respect to this process are equal to the
expectation values defined in the original system. This method fails in many cases, and it was not
until recently that with the help of new techniques \cite{Aarts:2009uq,Aarts:2011ax,Nishimura:2015pba,Nagata:2016vkn,Salcedo:2016kyy}
it was possible to meet the conditions for the  equivalence of the stochastic process 
defined by the CLM and the original system, and obtain correct results for an extended range of parameters
\cite{Seiler:2012wz,Nagata:2015uga,Tsutsui:2015tua,Ito:2016efb,Nagata:2016alq,Doi:2017gmk,Nagata:2018net}.
The CLM  has been applied successfully to many systems in lattice quantum
field theory 
\cite{Ambjorn:1985cv,Ambjorn:1986mf,Berges:2005yt,Berges:2006xc,Berges:2007nr,Pehlevan:2007eq,Aarts:2008rr,
      Aarts:2008wh,Aarts:2009hn,Aarts:2010gr,Aarts:2011zn,Seiler:2012wz,Sexty:2013ica,Hayata:2014kra,
      Fodor:2015doa,Aarts:2016qrv,Sinclair:2018rbk,
      Attanasio:2018rtq,Nagata:2018net,Nagata:2018mkb,Ito:2018jpo,Kogut:2019qmi,Sexty:2019vqx,Hirasawa:2020bnl},
and matrix models 
\cite{Mollgaard:2013qra,Mollgaard:2014mga,Ito:2016efb,Bloch:2017sex,Anagnostopoulos:2017gos,Nishimura:2019qal,
      Nagata:2018net,Basu:2018dtm,Joseph:2019sof,Anagnostopoulos:2019ptt,Anagnostopoulos:2020xai}.

In this talk we review the application of the CLM to the Euclidean
type IIB matrix model discussed in
\cite{Anagnostopoulos:2017gos,Anagnostopoulos:2019ptt,Anagnostopoulos:2020xai}.
The successful application of the method requires the deformation of
the model by two parameters $\mf$ and $\epsilonm$, which deform the
Dirac operator and the Bosonic part of the action with mass like
terms. This way, one can avoid the singular drift problem
\cite{Ito:2016efb}, which can lead to erroneous results in systems
where singularities of the drift dominate in the stochastic process of
the CLM \cite{Nishimura:2015pba}.  In the IIB matrix model the
singular drift problem occurs when the eigenvalues of the Pfaffian,
when put into a canonical Youla's form, accumulate near zero.  This
method was applied successfully on the 6D IIB model
\cite{Anagnostopoulos:2017gos}, obtaining SSB to SO($3$), consistent
with GEM calculations in \cite{Aoyama:2010ry}.  The 10D model was
studied in \cite{Anagnostopoulos:2020xai}. In that case, the
simulations were harder because the fermionic degrees of freedom
increase by a factor of four and the finite size effects are more
severe due to the increase of the dimensionality of the target space.
The results are consistent with GEM calculations \cite{Nishimura:2011xy},
showing SSB to  SO($3$). A careful extrapolation of the results, first
to large $N$, then to small $\epsilonm$ and finally to small $\mf$ is
necessary in order to obtain the correct large $N$ limit of the
original model. This makes the calculation tricky, and the Monte Carlo
simulations were done in parallel with GEM calculations of the
$\mf$--deformed model.  By computing the free energy of vacua of
different dimensionality using the GEM, it was possible to find
physical solutions for the SO($d$), $d=4,6,7$ ansatzes, calculated up
to three loops. As $\mf$ decreases, lower dimensional vacua turn out
to have lower free energies and the pattern of SSB is similar to
the one obtained by the CLM.

\section{The IKKT matrix model}
The IKKT matrix model  \cite{Ishibashi:1996xs} is defined by the action: \label{sec_IKKT}
\begin{eqnarray}
 S &=& S_{\textrm{b}} + S_{\textrm{f}} \ , \textrm{ where } 
\label{IKKT_action} \\
 S_{\textrm{b}} &=& - \frac{1}{4} N \, \textrm{tr} 
[A_{\mu}, A_{\nu}] [A^{\mu}, A^{\nu}] \ , 
\label{IKKT_boson} \\
 S_{\textrm{f}} &=& - \frac{1}{2} N \,  \textrm{tr} \left( 
\psi_{\alpha} 
({\cal C} \Gamma^{\mu})_{\alpha \beta} [A_{\mu}, \psi_{\beta}] \right) \ . 
\label{IKKT_fermion}
\end{eqnarray}
For $D=10$, the matrices $A_{\mu}$ ($\mu =0,1,2, \ldots, D-1$) are $N \times N$
traceless Hermitian matrices which transform like vectors, and
$\psi_{\alpha}$ ($\alpha =1,2, \ldots, 2^{D/2-1}$) are $N \times N$ traceless
matrices with Grassmann entries which transform like Majorana-Weyl spinors. 
The $2^{D/2-1} \times 2^{D/2-1}$ matrices $\Gamma^{\mu}$ and ${\cal C}$ are the 
gamma matrices after Weyl projection and the charge conjugation matrix, respectively, in ten dimensions.

In order to obtain the Euclidean version of the IKKT matrix model, we perform the Wick rotation
\begin{equation}
 A_{0} = i A_{D} \ , \quad \quad  \Gamma^{0} = - i \Gamma^{D} \  .
\label{wick_IKKT}
\end{equation}
The metric now is  $\delta_{\mu \nu}$ ($\mu,\nu=1,\ldots,D$) and the partition function becomes
\begin{equation}
 Z = \int dA d\psi\, e^{-S} = 
\int dA\, e^{-S_{\textrm{b}}}\, \textrm{Pf } {\cal M}\  ,
\label{EIKKT_partition}
\end{equation}
where the  $2^{D/2-1} (N^2-1) \times 2^{D/2-1}(N^2-1)$ anti-symmetric matrix ${\cal M}$ is defined by its action
\begin{equation}
 \psi_{\alpha} \to ({\cal M} \psi)_{\alpha} = 
({\cal C} \Gamma^{\mu})_{\alpha \beta} [A_{\mu}, \psi_{\beta}] 
\label{M_definition}
\end{equation}
on the linear space of traceless complex $N \times N$ matrices. The model has an  
SO($D$) rotational symmetry acting on $A_\mu$ and $\psi_\alpha$. 

Dynamical compactification of extra dimensions can be realized via 
the SSB of the SO($D$) symmetry to SO($d$), with $d<D$. The order parameters of the SSB
can be taken to be \cite{Ito:2016efb,Anagnostopoulos:2017gos}
\begin{equation}
\label{m:6}
\lambda_\mu = \frac{1}{N} \tr (A_\mu)^2 \ , \quad
\mu = 1,\ldots,D\  ,
\end{equation}
where no sum over $\mu$ is taken. Then, we break the  SO($D$) symmetry explicitly by deforming the
model by
\begin{equation}
\label{m:7}
\Delta S_{\rm b} = \frac{1}{2} N \epsilonm 
 \sum_{\mu=1}^{D} m_\mu  \tr\left(A_\mu\right)^2
\end{equation}
using the parameters $\epsilonm$ and  $0<m_1\leq\ldots\leq m_{D}$. The SSB pattern will 
arise in the $\epsilonm\to 0 $ limit {\it after} taking the large-$N$ limit. For finite
$N$, the choice  $0<m_1\leq\ldots\leq m_{D}$ yields $\vev{\lambda_1}\geq \ldots \geq \vev{\lambda_{D}}$.
If there is no SSB, all $\vev{\lambda_\mu}$ will be equal in the $N\rightarrow \infty$
and $\epsilon \rightarrow 0$ limits. Otherwise, SSB occurs and we will conclude that some
dimensions of space are larger than others.

The SSB of the SO($D$) rotational symmetry has been studied by using
the GEM. This method uses a systematic expansion around a
Gaussian action $S_0$, which contains many parameters.  In order to
make the calculation tractable, the number of parameters is reduced by
considering SO($d$)--symmetric ansatzes for $d<D$.  The free energy
and the expectation values of the observables is calculated in the
expansion around $S_0$ as a function of these parameters, and the
physical solutions are computed by finding a region in the parameter
space where the results are independent of these parameters.  In this
way, one obtains {\it nonperturbative} information about the model
\cite{Stevenson:1981vj}.  When applied to the IKKT matrix model
\cite{Nishimura:2001sx,Kawai:2002jk,Nishimura:2002va,Kawai:2002ub,Nishimura:2003gz,Nishimura:2004ts,Aoyama:2006di,Aoyama:2006rk,Aoyama:2006je,Aoyama:2010ry,Nishimura:2011xy},
the GEM yields the free energy and the average extent of spacetime in
each direction.  The $D=6$ case was studied in \cite{Aoyama:2010ry},
and free energy and the average extent of spacetime was calculated to
fifth order.  The SO($3$) ansatz was found to have the lowest free
energy, showing SSB to SO($3$).  The extended directions where found
to have an extent $R^2(d)$ for $\mu=1,\ldots,d$, whereas the shrunken
directions to have and extent $r^2=\vev{\lambda_\mu}\approx 0.223$,
$\mu=d+1,\ldots,D$, which is independent of $d$. The values of
$R^2(d)$ are such that
\begin{equation}
\label{e02}
R^d(d)\, r^{D-d} \approx l^D\, ,
\end{equation}
where $v\equiv l^D$ is the spacetime volume. It turns out that
$l^2\approx 0.627$ and $R^2(3)\approx 1.76$.  The $D=10$ case was
studied in \cite{Nishimura:2011xy}, where a systematic expansion to
third order was performed for SO($d$) ansatzes with $2\leq d \leq
7$. The SO($3$) ansatz turns out to have the lowest free energy,
also showing SSB to SO($3$).  In that case
\begin{equation}
\label{e03}
R^2 (3) \approx 3.27 \ , \quad\quad
r^2 \approx 0.155 \ , \quad\quad
l^2 \approx 0.383 \ .
\end{equation}
These results are consistent with the ones obtained by Monte Carlo simulations \cite{Anagnostopoulos:2015gua}.
They are finite, in contrast to the case of the Lorentzian model, where $R^2(d)$ expands indefinitely.

\section{The Complex Langevin Method}

In this section we review how to apply the CLM to the Euclidean IKKT matrix model.
The partition function can be written as
\begin{equation}
Z = \int dA\,{\rm e}^{-\Seff} \ ,
\label{m:1}
\end{equation}
where the effective action $\Seff=S_{\rm b}-\log\Pf{\cal M}$ is complex. In the CLM the matrices $A_\mu$
become general complex traceless matrices, which amounts to the complexification of the degrees of
freedom in the CLM. The time evolution in the fictitious Langevin time is given by
\begin{equation}
\label{m:2}
\frac{d\left(A_\mu(t)\right)_{ij}}{dt} = 
-\frac{\partial \Seff[A_\mu(t)]}{\partial \left(A_\mu\right)_{ji}}
+\left(\eta_\mu\right)_{ij}(t) \  .
\end{equation}
The noise  $\eta_\mu(t)$ is made from traceless Hermitian matrices whose elements are random variables 
obeying the Gaussian distribution 
$\propto\exp\left(-\dfrac{1}{4}\int\tr\left\{\eta_\mu(t)\right\}^2\,dt\right)$. 
The first term on the right-hand side is the drift term
\begin{equation}
\label{m:3}
 \frac{\partial\Seff    }{\partial \left(A_\mu\right)_{ji}}
=\frac{\partial S_{\rm b}     }{\partial \left(A_\mu\right)_{ji}}
-\frac{1}{2}\Tr\left(
{\cal M}^{-1} \frac{\partial {\cal M}}{\partial \left(A_\mu\right)_{ji}}
\right)\  ,
\end{equation}
where $\Tr$ represents the trace of a $16(N^2-1)\times 16(N^2-1)$
matrix. 
The expectation value of an observable ${\cal O}[A_\mu]$ 
can be calculated from
\begin{equation}
\label{m:4}
\vev{{\cal O}[A_\mu]} = \frac{1}{T}\int_{t_0}^{t_0+T}{\cal O}[A_\mu(t)]dt\  ,
\end{equation}
where $A_\mu(t)$ is a general complex matrix 
solution of \rf{m:2}, $t_0$ is the thermalization time, 
and $T$ is large enough in order to obtain good statistics. 
Upon complexification of the matrices $A_\mu(t)$, 
the observable ${\cal O}[A_\mu(t)]$
depends on general complex matrices. 
The analyticity of the function
${\cal O}[A_\mu]$ plays a crucial role in the proof of the
validity of \rf{m:4}\cite{Aarts:2009uq,Aarts:2011ax,Nagata:2016vkn}.
The discretization of  \rf{m:2} is given by
\begin{equation}
\label{m:5}
\left(A_\mu\right)_{ij}(t+\Delta t) = \left(A_\mu\right)_{ij}(t)
-\Delta t  \frac{\partial S[A_\mu(t)]}{\partial \left(A_\mu\right)_{ji}}
+\sqrt{\Delta t}\, \left(\eta_\mu\right)_{ij}(t)\  ,
\end{equation}
which we have used in order  to solve \rf{m:2} numerically. 

The solutions of the stochastic equation \rf{m:2} are random variables with distribution
$P(A_\mu^{({\rm R})},A_\mu^{({\rm I})};t) $, where 
$A_\mu^{({\rm R})}(t)=(A_\mu(t)+A_\mu^\dagger(t))/2$, 
$A_\mu^{({\rm I})}(t)=(A_\mu(t)-A_\mu^\dagger(t))/2i$, and in order that in the $t\to\infty$ limit an 
observable ${\cal O}[A_\mu]$ has an expectation value in this distribution equal to
$\vev{{\cal O}[A_\mu]}$, given by the path integral \rf{EIKKT_partition}, certain 
conditions need to be met. In \cite{Nagata:2016vkn} it was shown that if the magnitude of the drift
falls off exponentially or faster, then these conditions are met. For this, the $A_\mu(t)$ should not
make long trips in the anti-Hermitian direction. Gauge cooling has been applied in order to restrict 
those trips and meet this condition\cite{Anagnostopoulos:2017gos,Anagnostopoulos:2020xai}. Adaptive 
stepsize techniques have also been applied in order to keep the stability of the time evolution.
Furthermore, one has to avoid the singular drift problem. This problem occurs in the term with 
${\cal M}^{-1}$ in Eq.~\rf{m:3},
when the eigenvalues of $\cal M$ accumulate densely near zero. 
In order to avoid this problem, we deform the fermionic action by adding a term
\begin{equation}
\label{m:11}
\Delta S_{\rm f} = -i \mf \frac{N}{2} \, \tr\left(
\psi_\alpha 
({\cal C}\Gamma_8 \Gamma_9^\dagger \Gamma_{10})_{\alpha\beta}
\psi_\beta
\right)\  ,
\end{equation}
where $\mf$ is the deformation parameter. This term shifts the
eigenvalue distribution of $\cal M$ away from the origin
\cite{Ito:2016efb}, but it also breaks the SO($10$) symmetry down to
${\rm SO}(7)\times {\rm SO}(3)$ explicitly.  Therefore, we examine if
the remaining SO($7$) symmetry breaks down to smaller subgroups as
$\mf$ is varied and discuss what occurs at $\mf = 0$. As
$\mf\to\infty$, the fermionic degrees of freedom decouple and we
obtained the so--called bosonic model. This model is known to be
SO($10$) symmetric and no SSB occurs \cite{Anagnostopoulos:2015gua}.

The severeness of the singular drift problem depends on the parameters
$\mf$ and $\epsilonm$. For large enough $\mf$ the problem disappears,
but as $\mf$ is lowered the problem reappears for small enough
$\epsilonm$.  In our simulations we monitor the fall off of the drift
and make sure that it falls of faster than exponentially. The singular
drift problem is expected to vanish for large enough $N$
\cite{Anagnostopoulos:2020xai}.

\section{Results}

In our numerical investigation we consider whether the remaining
SO($7$) symmetry is broken down to a smaller group.  We study the
model deformed by \rf{m:7} and \rf{m:11}. We use $\mf = 3.0$, $1.4$, $1.0$, $0.9$, $0.7$,
in order to extrapolate to the undeformed IKKT model at $\mf = 0$.
The $m_\mu$ in Eq.~\rf{m:7} are chosen so that this term does not
break SO($10$) completely, because otherwise the spectrum of $m_\mu$
becomes too wide to make the $\epsilonm\to 0$ extrapolation reliably.
For $\mf=3.0$, we choose $m_\mu=(0.5, 0.5, 0.5, 1, 2, 4, 8, 8, 8, 8)$,
which enables us to distinguish SO($d$) vacua with $d=3,4,5,6,7$.  For
smaller values of $\mf$, we choose $m_\mu=(0.5, 0.5, 1, 2, 4, 8, 8, 8,8,8)$, 
which enables us to distinguish SO($d$) vacua with $d=2,3,4,7$.
In particular, we may confirm that the SO(3) symmetry remains unbroken
by seeing that $\langle \lambda_1 \rangle = \langle \lambda_2 \rangle$
and $ \langle \lambda_3 \rangle$ agree in the $N \rightarrow \infty$
and $\epsilonm \rightarrow 0$ limits.  On the other hand, this choice
of $m_\mu$ has a drawback that $\lambda_6$ and $\lambda_7$ are mixed
up because of $m_6 = m_7$, and hence one cannot distinguish SO(5) and
SO(6) vacua.  This does not cause any harm, however, as far as we find
that $\langle \lambda_4 \rangle$ and $\langle \lambda_5 \rangle$ do
not agree in the $N \rightarrow \infty$ and $\epsilonm \rightarrow 0$
limits, which implies that the SO(7) symmetry is broken to SO(4) or
lower symmetries.

\begin{figure}[htbp]
\centering 
\includegraphics[width=0.45\textwidth]{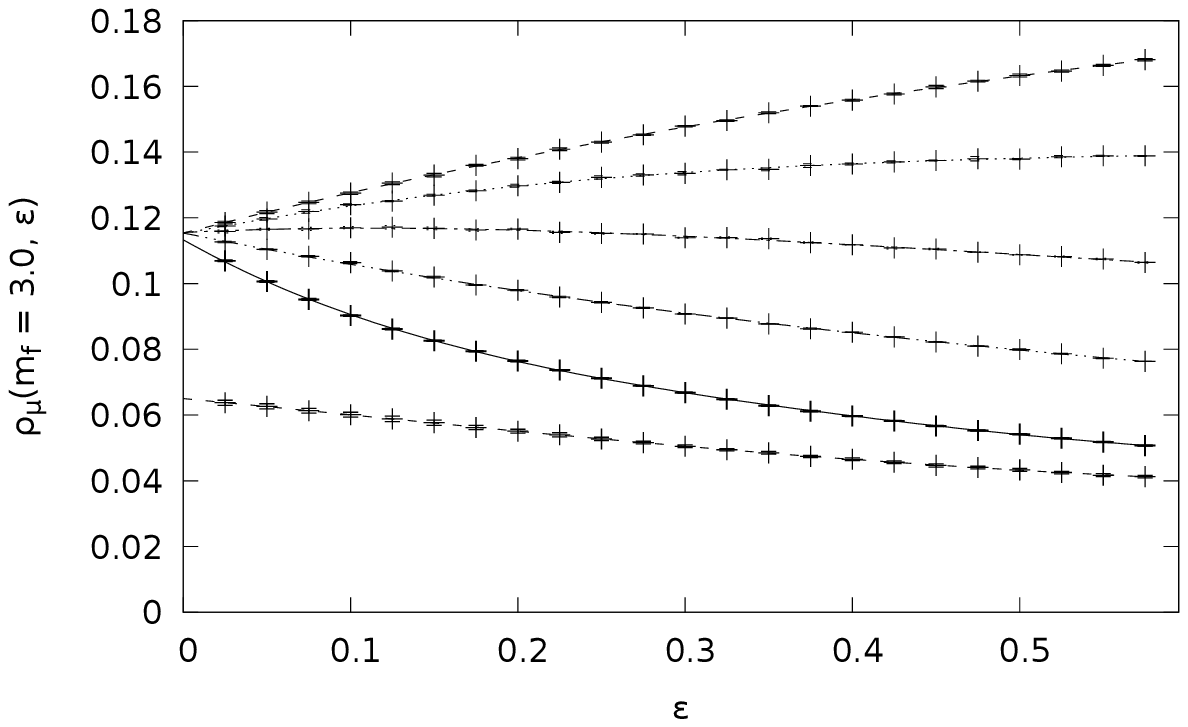} 
\includegraphics[width=0.45\textwidth]{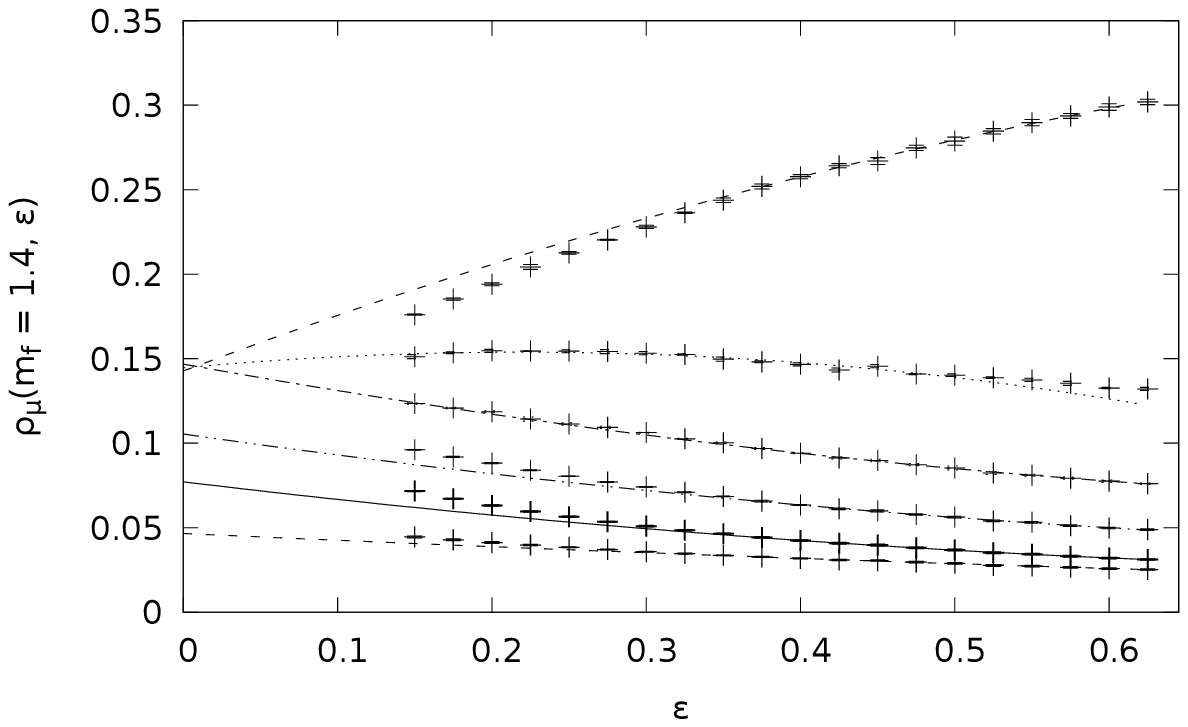}
\includegraphics[width=0.45\textwidth]{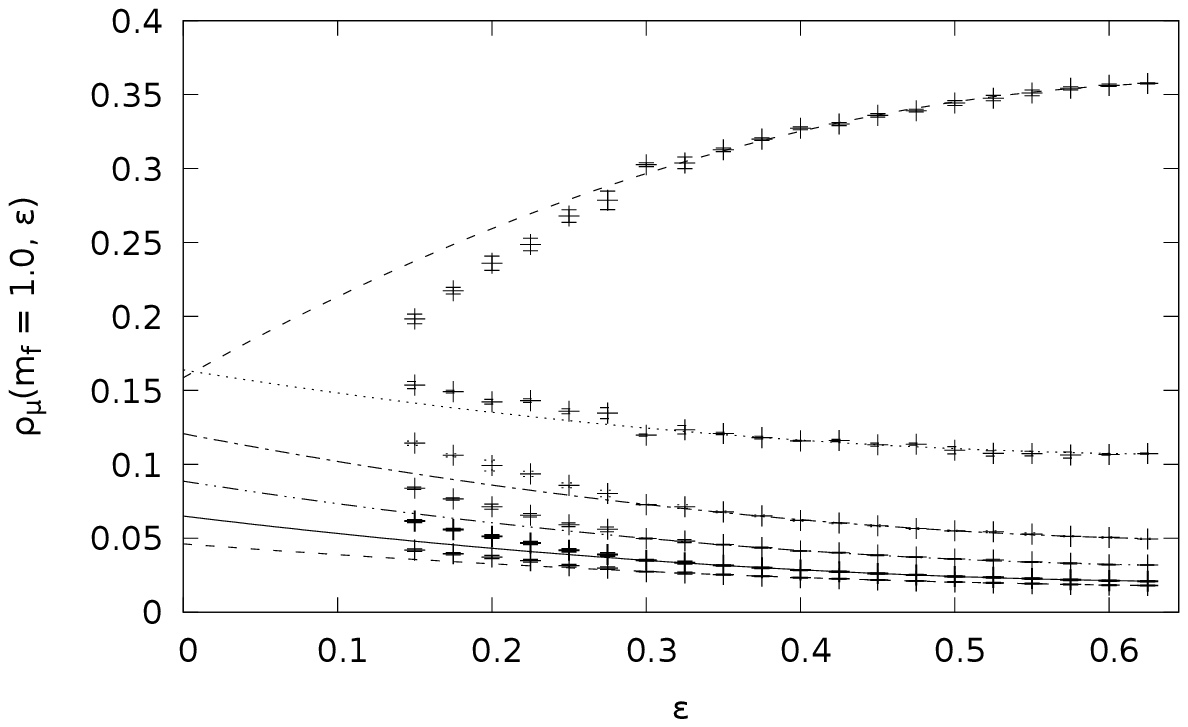}
\includegraphics[width=0.45\textwidth]{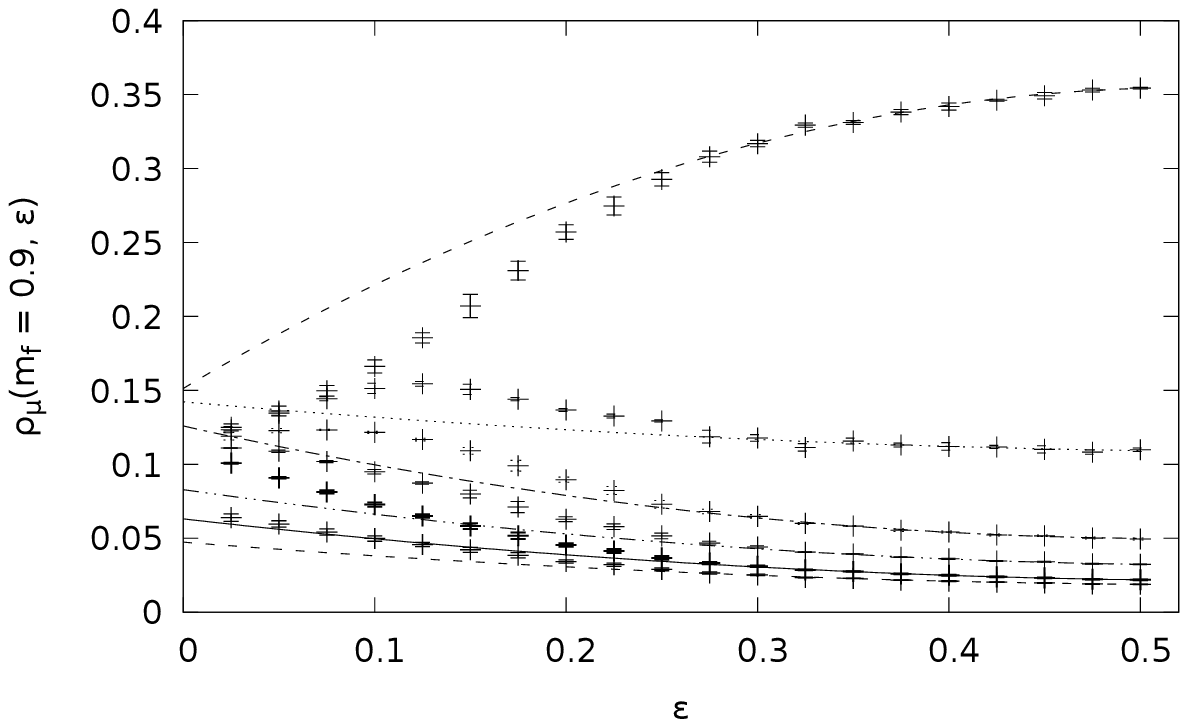}
\includegraphics[width=0.45\textwidth]{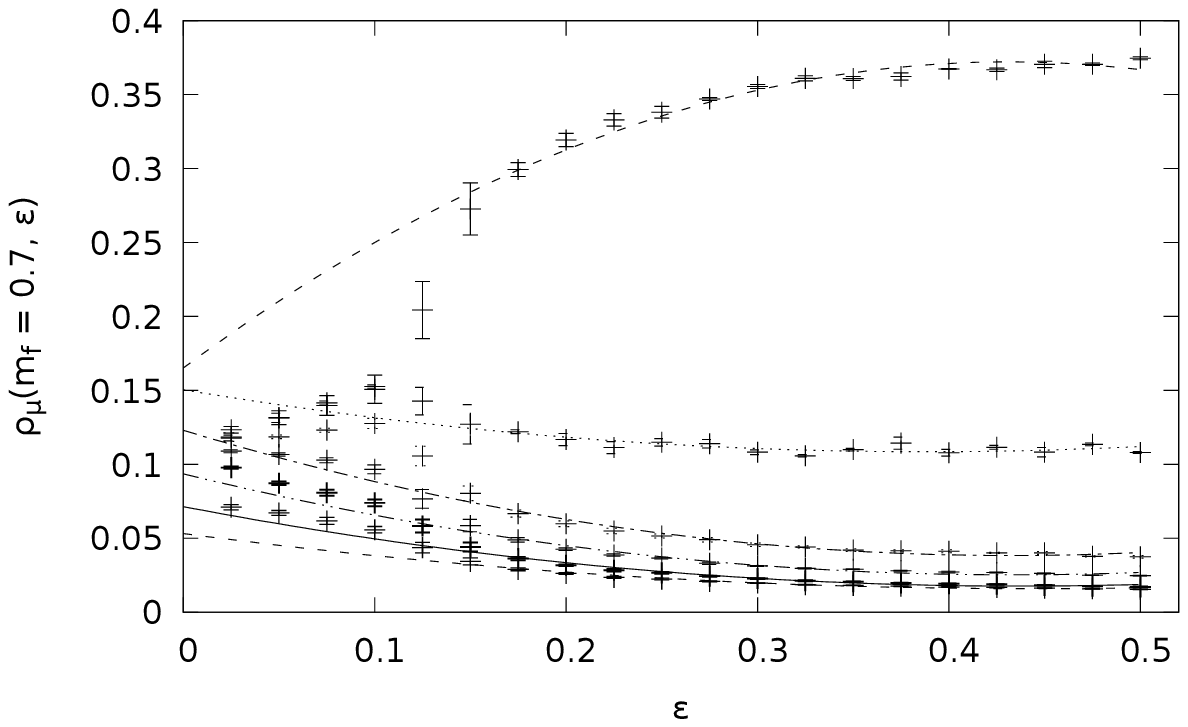}
    \caption{The $\rho_\mu(\mf,\epsilonm)$ in Eq.~\protect\rf{r02} 
are plotted against $\epsilonm$ for $\mf=3.0$ (Top-Left), 
$\mf=1.4$ (Top-Right), $\mf=1.0$ (Middle-Left),
$\mf=0.9$ (Middle-Right) and $\mf=0.7$ (Bottom). 
We use
$m_\mu=(0.5, 0.5, 0.5, 1, 2, 4, 8, 8, 8, 8)$
for $\mf=3.0$ and $m_\mu=(0.5, 0.5, 1, 2, 4, 8, 8, 8, 8, 8)$
for the other values of $\mf$.
The continuous lines are polynomial fits in $\epsilonm$. 
For $\mf=3.0$ a quartic fit is performed, 
whereas for the other values of $\mf$ the
fits are quadratic in $\epsilonm$. 
In the $\mf=3.0$ plot, the curves from top to bottom are 
$(\rho_1+\rho_2+\rho_3)/3$, $\rho_4$, $\rho_5$, 
$\rho_6$, $\rho_7$  and $(\rho_8+\rho_9+\rho_{10})/3$. 
For the other plots, the curves from top to bottom are 
$(\rho_1+\rho_2)/2$,  $\rho_3$, $\rho_4$, $\rho_5$, $(\rho_6+\rho_7)/2$  
and $(\rho_8+\rho_9+\rho_{10})/3$.
\label{f:r02}}
\end{figure}

In order to probe the SSB, one has to take the $N\to\infty$ limit
first and then the $\epsilonm\to 0$ limit.  For given $m_\mu$ in
Eq.~\rf{m:7}, the large-$N$ limit is obtained by first computing the
ratio
\begin{equation}
\label{r01}
 \rho_\mu(\mf, \epsilonm,N) = 
  \frac{\vev{\lambda_\mu}_{\mf, \epsilonm,N}}
{\sum_{\nu=1}^{10}\vev{\lambda_\nu}_{\mf, \epsilonm,N}}\  ,
\end{equation}
and then by making a large-$N$ extrapolation
\begin{equation}
\label{r02}
 \rho_\mu(\mf, \epsilonm)= \lim_{N\to \infty} \rho_\mu(\mf, \epsilonm,N)\ .
\end{equation}
The large-$N$ extrapolation is performed by plotting $\rho_\mu(\mf,\epsilonm,N)$ 
against $1/N$ and making a quadratic fit with respect to $1/N$. Then we make the $\epsilonm\to 0$ extrapolation
\begin{equation}
\label{r03}
\rho_\mu(\mf) = \lim_{\epsilonm\to 0} \rho_\mu(\mf,\epsilonm)
\end{equation}
by fitting $\rho_\mu(\mf,\epsilonm)$ to a polynomial in $\epsilonm$.
In Fig.~\ref{f:r02}
we plot the large-$N$ extrapolated values 
$\rho_\mu(\mf,\epsilonm)$ as a function of $\epsilonm$ for
$\mf=3.0$, 1.4, 1.0, 0.9 and $0.7$ together with the performed fits.
In those fits, the data that is affected by finite size effects must be excluded.
As it was argued in \cite{Anagnostopoulos:2020xai}, these become quite severe in the $\epsilonm\to 0$ limit.
These effects are apparent in Fig.~\ref{f:r02}, where a crossover to a symmetric
phase is observed for $\mf\le 1.4$ as  $\epsilonm\to 0$. This crossover is expected to 
vanish in the large--$N$ limit.

From the extrapolated values $\rho_\mu(\mf)$, we find that the SO($7$)
symmetry of the deformed model is not spontaneously broken at $\mf =
3.0$, but it is actually broken to SO($4$) for $\mf = 1.4$ and to
SO($3$) for $\mf = 1.0$, $0.9$, $0.7$.  Thus as $\mf$ is decreased,
the SO($7$) symmetry seems to be spontaneously broken to smaller
subgroups gradually in the same way as it was observed in the $D=6$
case \cite{Anagnostopoulos:2017gos}.  However, we consider that the
symmetry is not going to be broken further down to SO($2$) at smaller
$\mf$.  This is based on the fact that the Pfaffian vanishes
identically for strictly 2D configurations
\cite{Nishimura:2000ds,Nishimura:2000wf}, which implies that the
mechanism of SSB due to the phase of the Pfaffian no longer works
there\footnote{This is also reflected in the GEM results
  \cite{Aoyama:2010ry,Nishimura:2011xy} for the free energy of the
  SO($d$) vacuum, which becomes much larger for $d=2$ than for $d\ge
  3$.}.  Hence our results are consistent with the results obtained by
the GEM for the undeformed model, which show that the SO(3) vacuum has
the smallest free energy.

\begin{figure}
\centering{}\includegraphics[width=0.6\textwidth]{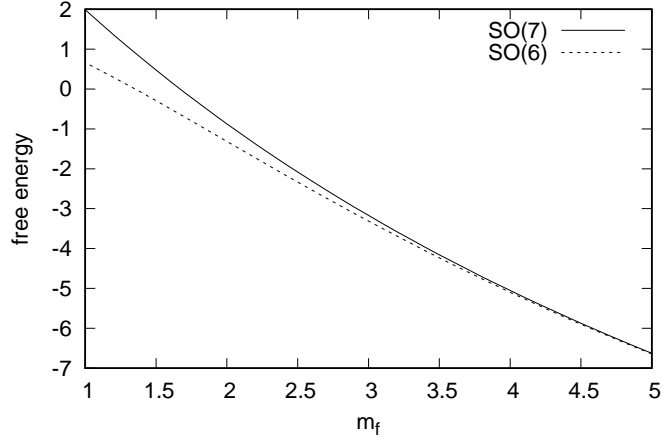}
\caption{The free energy calculated up to three loops for the 
solutions found with the SO($7$) and SO($6$) ansatzes
are plotted against the fermion mass $\mf$. 
We observe a clear tendency that the SO(6) symmetric vacuum is more favored
as $\mf$ is decreased, whereas the free energy for the two ansatzes
tends to be degenerate as $\mf$ is increased.
\label{fig:free_energy_gem}}
\end{figure}

The above mentioned results have also been checked against a
calculation using the GEM on the deformed model \rf{m:11}. The GEM has
been applied to the Euclidean IKKT matrix model in
\cite{Nishimura:2011xy}, where SO($10$) was found to be broken down to
SO($3$).  The GEM has the advantage that the large-$N$
limit is easily taken by considering only planar graphs and that
the small-$\epsilonm$ extrapolation is not necessary. The systematic errors
of the GEM are due to the truncation of the expansion and the errors
in determining the parameters that give the physical solutions. As
such, the two methods can be considered to be completely independent.
In \cite{Anagnostopoulos:2020xai}, we performed a three-loop
calculation using SO($d$) symmetric ansatzes for $d=6, 7$, and
calculated the free energy. We observed that by decreasing $\mf$, the
free energy of the SO($6$) vacuum becomes smaller than the free energy
of the SO($7$) vacuum. In Fig.~\ref{fig:free_energy_gem} we plot the
free energy calculated up to three loops for the solutions found with
the SO($7$) and SO($6$) ansatzes against the fermion mass $\mf$.  We
observe a clear tendency that the SO(6) symmetric vacuum is more
favored as $\mf$ is decreased.  However, the free energy for the two
ansatzes tends to become degenerate as $\mf$ is increased.  In this
situation it is difficult to identify the critical point, given the
accuracy of the GEM results.

At $\mf=3.0$, the extent of space was found to agree very well between the
two methods. In \cite{Anagnostopoulos:2020xai}, it is found that
\begin{equation}
\label{kk:01}
\rho_1 = \cdots = \rho_7 = 0.116  \ , \quad
\rho_8 = \rho_9 = \rho_{10} = 0.064\, ,
\end{equation}
whereas the CLM results of Fig.~\ref{f:r02} give
\begin{eqnarray}
\rho_1 = \cdots = \rho_7 = 0.115  \ , \quad
(\rho_8 + \rho_9 + \rho_{10})/3 = 0.065  \ .
\label{extent_CLM}
\end{eqnarray}

Therefore, for the first time, we have a first principle study of the
Euclidean IKKT matrix model that produced clear results on the
question of dynamical compactification of extra dimensions via SSB of
the SO($10$) rotational symmetry of the model. The SO($10$) rotational
symmetry breaking of the Euclidean IKKT matrix model down to SO($3$) due to the
phase of the Pfaffian is interesting, but it makes the model somewhat
difficult to interpret.  Given the promising properties of the
Lorentzian model \cite{Kim:2011cr,Ito:2017rcr,Ito:2013ywa,Ito:2013qga,Ito:2015mem,Ito:2015mxa,Azuma:2017dcb,
  Aoki:2019tby,Nishimura:2019qal}, we consider that
the naive Wick rotation to the Euclidean model is not the right
direction to pursue. On the other hand, the fact that the CLM enabled
us to obtain a clear SSB pattern for the deformed model, which suffers
from a severe sign problem, is encouraging.  We hope that the CLM is
equally useful in investigating the Lorentzian IKKT model, in
particular in the presence of fermionic matrices, which are not
included yet in Ref.~\cite{Nishimura:2019qal}.

\section*{Acknowledgements}
The authors would like to thank S.~Iso, H.~Kawai, H.~Steinacker and
A.~Tsuchiya for valuable discussions.  T.~A.\ was supported in part by
Grant-in-Aid for Scientific Research (No.\ 17K05425) from Japan
Society for the Promotion of Science.  Computations were carried out
using computational facilities at KEKCC and the NTUA het cluster.
This work was also supported by computational time granted by the
Greek Research \& Technology Network (GRNET) in the National HPC
facility - ARIS - under project ID IKKT10D.

\bibliographystyle{JHEP}
\bibliography{ref}

\end{document}